\documentclass[12pt]{article}
\usepackage{a4,epsfig,amsmath,amssymb,cite}

\begin{document}

\begin{center}
   \centering{\large\bf Heat capacity of protein folding}\\

\vspace*{1cm}
\centering{Audun Bakk\footnote{Corresponding author. Electronic address: Audun.Bakk@phys.ntnu.no}, Johan S. H\o ye\footnote{Electronic address: Johan.Hoye@phys.ntnu.no}, and Alex Hansen\footnote{Electronic address: Alex.Hansen@phys.ntnu.no}\\
           {\it Department of Physics, Norwegian University 
                of Science and  Technology, 
                NTNU, NO--7491 Trondheim, Norway}}
  \vspace*{0.5cm}

  \centering{(\today)}
\end{center}
\vspace*{0.5cm}

%*************************************************************
% ABSTRACT
%*************************************************************
\begin{abstract}
  We construct a Hamiltonian for a single domain protein where the contact enthalpy and the chain entropy decrease linearly with the number of native contacts. The hydration effect upon protein unfolding is included by modeling water as ideal dipoles that are ordered around the unfolded surfaces, where the influence of these surfaces, covered with an ``ice-like'' shell of water, is represented by an effective field that directs the water dipoles. An intermolecular pair interaction between water molecules is also introduced. The heat capacity of the model exhibits the common feature of small globular proteins, two peaks corresponding to cold and warm unfolding, respectively. By introducing vibrational modes, we obtain quantitatively good accordance with experiments.

\end{abstract}


%*************************************************************
% MAIN TEXT
%*************************************************************
\section{Introduction}
\label{sec:1}
A protein is a large polymer consisting of many thousands of atoms, and may therefore from a physical point of view be regarded as a macroscopic system\cite{Creighton:92a}. Anfinsen\cite{Anfinsen:73} proved that the one-dimensional sequence of amino residues determines uniquely the three-dimensional conformation of the protein. Further he concluded that the native (folded) state is the state of the lowest free energy. Consequently, given a polypeptide sequence, a microscopic analysis of its enthalpy, and degrees of freedom, followed by a statistical mechanical evaluation should reveal its thermodynamical properties of a given protein. 

Proteins are known to fold on time scales from milliseconds to seconds, in great contrast to a statistical analysis based on the vast number of conformations in the energy landscape which indicates an astronomical folding time. One way to circumvent this paradox, attributed to Levinthal\cite{Levinthal:68}, is to suppose some kind of folding pathway\cite{Baldwin:99,Wang:95,Hansen:98a,Hansen:99,Hansen:98b,Bakk:00a,Bakk:00b}. The latter means that the folding process, starting from a denaturated (unfolded) conformation, follows a specific sequence of folding steps in the free energy landscape, until the native state is reached. 

Proteins consist of 20 different amino acids with a great diversity with regard to size, polarity and charge. Considering the electrostatics it is possible to argue that the small number of charges does not contribute significantly to the stability of the native conformation\cite{Creighton:92b}. Thus, two kinds of surfaces remain most relevant, the apolar (hydrophobic) and polar surfaces\cite{Creighton:92a}.

In this work we model a small single-domain protein consisting of energetically equal contacts that lower their energy when they are closed, and it decreases linearly with the number of ``native like'' contacts. In addition, we introduce water molecules, modeled as ideal electrical dipoles in an external effective field that represent the influence or ordering effect of the unfolded apolar surfaces on the water. Besides, there are pair interactions between water molecules. In a self-consistent mean field treatment they add to the effective field\cite{Hoye:80,Ma:85}. By performing statistical mechanical evaluations we then find thermodynamic functions like the free energy and the heat capacity of the protein. In the end we assign vibrational modes in the IR-region, and compare the heat capacity to experimental results on metmyoglobin\cite{Privalov:86}.

%--------------------------------------------------------------------
%--------------------------------------------------------------------
\section{The protein model}
\label{sec:2}
In this section we will present the statistical model which is a further development of earlier models of Hansen {\it et al.}\cite{Hansen:98a,Hansen:99} and Bakk {\it et al.}\cite{Bakk:00a,Bakk:00b}. The chain-chain interactions, although reformulated, are used as in these models, but the protein-water interactions are new compared to these models. Thus, we will spend the greater part of this section discussing protein-water and water-water interactions. 

%--------------------------------------------------------------------
\subsection{Internal forces in the protein}
\label{sub:21}
For simplicity the protein is regarded as consisting of $N$ {\it contacts}\cite{Plaxco:98}. A contact is a conformation with a specified free energy. Beyond this specification the model does not have a detailed connection to the structure of proteins. However, the general character of the model makes it possible to reveal various key features about the specific mechanism of protein folding thermodynamics that can lead to cold and warm unfolding. Each contact is supposed to be a specific point on a folding pathway\cite{Baldwin:99,Wang:95,Hansen:98a,Hansen:99,Hansen:98b,Bakk:00a,Bakk:00b}. The pathway is also in accordance with a ``folding funnel''\cite{Leopold:92}, where each contact now represents the intersection between a level or a ``contour line'' in the free energy landscape and one of the possible multiple pathways\cite{Galzitskaya:99}. 

Upon folding, the protein lowers its enthalpy by an amount $\epsilon_{\rm c}$ for each contact that ``folds''\cite{Bryngelson:87}. Let $i\in \{0,1,...\,,N\}$ be the number of contacts that are correctly folded. Thus the resulting energy  for the bare or vacuum chain-chain interactions can be written
\begin{equation}
   \label{eq:H_chain}
   {\cal H}_i^{\rm c}=-i\,\epsilon_{\rm c}.
\end{equation}
The specific value $i=0$ means an unfolded (denaturated) protein, while $i=N$ is a complete folded (native) protein. To incorporate the rotational freedom or flexibility in the polypeptide backbone, we assign $g_{\rm c}$ degrees of freedom for each unfolded contact, where $g_{\rm c}$ is interpreted as the relative increase of the degrees of freedom for an unfolded contact compared to a folded one. Consequently the folding of $i$ contacts corresponds to $g_{\rm c}^{N-i}$ degrees of freedom. It is worth noting that this chain-chain interaction model can be viewed as a further simplification of the model by Zwanzig\cite{Zwanzig:95}.

%--------------------------------------------------------------------
\subsection{Hydration effect upon unfolding} 
\label{sub:22}
It is known from experiments that the heat capacity change upon aqueous dissolution of apolar molecules from their gaseous state is positive and proportional to the solute molecule concentration\cite{Edsall:35,Privalov:92}, and this change decreases with increasing temperature\cite{Privalov:88}. Further a solution of an apolar substance in water is associated with a {\it negative} entropy change at room temperature, which decreases in absolute value with increasing temperature\cite{Creighton:92a}. In other words, there seems to be an ordering of the water around the apolar surfaces. In sum the hydration effect of an apolar molecule can be explained by a gradual melting of an ordered ``ice-like'' shell around these compounds\cite{Frank:45}. Melting of ice is a complex process whereupon conformational changes implies a change in the enthalpy and the entropy. In this paper we incorporate the hydration of the apolar surfaces, by an extension of a model proposed by Hansen {\it et al.}\cite{Hansen:98a,Bakk:00c}.
 
The idea with the water interactions is to cover two basic properties of the model. First, there is an ordering of water around unfolded parts of the protein. This is accompanied by decreased enthalpy and entropy upon hydration of apolar molecules. Secondly, there are interactions between the water molecules. These interactions tend to orient the molecules with respect to each other to form an ``ice-like'' structure. The water molecules form hydrogen bonds at tetrahedral angles with neighboring water molecules. These bonds are associated with location of positive and negative charges within the water molecules. This again results in large permanent electric dipole moments of these molecules. Thus, we here as a simplification, approximate the water molecules by ideal electric dipoles. Apolar surfaces, in combination with hydrogen bonds, make it favorable for the water to make ``ice-like'' shell structures around these surfaces. The influence of these apolar surfaces we thus will model by an electric field ${\bf E}$, that also has a structuring effect as it rectifies the dipole moments. This field yields an interaction for each dipole
\begin{equation}
   \label{eq:H_E}
   {\cal H}_{\rm E}=-{\bf E}\cdot{\bf s}=-E\cos{\vartheta}.
\end{equation}
Here is ${\bf s}$ the dipole moment of the molecule where we for simplicity put $|{\bf s}|=1$, and $\vartheta$ is the angle between ${\bf E}$ and ${\bf s}$. 

Besides, there will be pair interactions between neighboring molecules with the total energy 
\begin{equation}
   \label{eq:H_p}
   {\cal H}_{\rm p}=-\frac{1}{2}\sum_{i,j}J_{ij}\,{\bf s}_i\cdot{\bf s}_j,
\end{equation}
 where $J_{ij}$ is the coupling constants between water molecule $i$ and neighboring molecules $j$. The factor $1/2$ prevents double counting of interactions. In our model these can be regarded as interactions between the water dipole moments. We find reasons to take such interactions into account as formation of ice also represents a directional ordering of water molecules, and we want to investigate their influence. In Appendix we calculate the partition function for one water molecule, $Z_{\rm w}$ in Eq.\ \ref{eq:Z_w}, by a mean field approximation\cite{Hoye:80,Ma:85} where the field $E$ is replaced by an effective field $E_{\rm e}=E+bm$.

The resulting term $i$ in the canonical partition function for the protein has $i$ contact energies $\epsilon_{\rm c}$, and $g_{\rm c}$ degrees of freedom together with $M$ ``bound'' water molecules each contributing a factor $Z_{\rm w}$ at each of the $N-i$ unfolded contacts, thus
\begin{equation}
   \label{eq:Z_i}
   Z_i=g_c^{N-i}e^{i\beta\epsilon_{\rm c}}(4\pi)^{Mi}
       \left(Z_{\rm w}\right)^{M(N-i)},
\end{equation}
where the factor $4\pi$ is the $Z_{\rm w}$ for $E=0$ for the $Mi$ ``unbound'' bulk water molecules.

Eq.\ \ref{eq:Z_i} further can be rewritten as 
\begin{equation}
   \label{eq:Z_i2}
   Z_i=e^{\beta N\epsilon_{\rm c}}(4\pi)^{MN}r^{i-N} ,
\end{equation}
where the function $r$, when inserting the $Z_{\rm w}$ from Appendix (Eq.\ \ref{eq:Z_w}), is 
\begin{equation}
   \label{eq:r}
   r\equiv\left[ a\,e^{\beta\mu}\,e^{\frac{1}{2}\beta bm^2}
                 \frac{\beta E_{\rm e}}
                      {\sinh{\beta E_{\rm e}}}\right]^M \quad ,
\end{equation}
with $a\equiv 1/g_c^{1/M}$, and $\mu\equiv\epsilon_c/M$. The parameters $a$ and $\mu$ will depend upon the chemical environments (pH, denaturant concentration, etc.\@) 

The canonical partition function for the system is now simply the sum over $Z_i$ for the various contact conformations along the folding pathway 
\begin{equation}
   \label{eq:Z}
   Z=\sum_{i=0}^N Z_i=e^{\beta N\epsilon_{\rm c}}(4\pi)^{MN}
                      \frac{1-r^{-(N+1)}}{1-r^{-1}}\quad .
\end{equation}
From the definition of the internal energy, $U=-\partial (\ln{Z})/\partial\beta$, it is clear from Eq.\,\ref{eq:Z} that the exponential contributes with a constant factor to $U$, thus the heat capacity, $C=-\beta^2\partial U/\partial\beta$, will only depend on the function $r$ in Eq.\,\ref{eq:r}. 

The number of {\it effective} parameters is therefore, for a fixed system size ($N=100$ in this work), reduced to only four: $a$, $b$, $\mu$, and $M$, as $E$ in Eq.\,\ref{eq:H_E} can be included in $\beta$, i.\@e., one may put $E=1$, as we have done.

%--------------------------------------------------------------------
\subsection{Vibrational modes} 
\label{sub:23}
As the results will show, the model above yields a heat capacity that lacks certain features. Experimentally the heat capacity by hydration increases markedly with $T$, and the ``valley'' in the folded region is well above zero. Thus to account for these latter features we will introduce vibrational modes. These are internal modes of the protein due to couplings between neighboring atoms, and they can be considered as harmonic oscillators. The quantization of the latter yield the energy levels
\begin{equation}
   \label{eq:H_h}
   {\cal H}_{\rm h}(n)=(n+\frac{1}{2})h\nu,
\end{equation}  
where $h=6.63\cdot 10^{-34}$ Js is  Planck's constant and $\nu$ is the frequency. Summing over all energy levels gives us the partition function for the vibrational modes for $N_{\rm h}$ independent harmonic oscillators
\begin{equation}
   \label{eq:Z_h}
   Z_{\rm h}=\left(\sum_{n=0}^{\infty}
       e^{-\beta{\cal H}_{\rm h}(n)}\right)^{N_{\rm h}}
      =\left(2\sinh{(d/T)}\right)^{-N_{\rm h}},
\end{equation}
where $d\equiv h\nu/(2k_{\rm B})$. We suppose, as a very simple assumption, that the vibrational modes are independent of the degree of folding, thus the partition function for the system including these,  is
\begin{equation}
   \label{eq:Z'}
   Z'=Z_{\rm h}Z,
\end{equation} 
where $Z$ is the partition function in Eq.\ \ref{eq:Z}. 
%-------------------------------------------------------------------
\section{Results and discussion}
\label{sec:3}
Whether the protein is folded or unfolded can now be analyzed in a straight-forward way by regarding the ratio $r=Z_{i+1}/Z_i\,$. The contribution $Z_i$ to the full partition function $Z$ may be regarded as the partition function for a protein with $i$ contacts folded. Thus, a free energy difference $\Delta_{\rm n}^{\rm d}F$ between the denaturated (d) and the native (n) protein can be expressed as
\begin{equation}
   \label{eq:delta_f}
   \Delta_{\rm n}^{\rm d}F=-T(\ln{Z_0}-\ln{Z_N})=TN\ln{r}\quad ,
\end{equation}
which determines the stable conformation, and gives a direct interpretation of the function $r$. For $\Delta_{\rm n}^{\rm d} F>0$ or $r>1$ the native conformation is thermodynamically stable, while for $\Delta_{\rm n}^{\rm d} F<0$ or $r<1$ the denaturated conformation is stable. The $r=1$ is a critical value, and in a small region around this value the protein switches between the two conformations. 

 In Fig.\,\ref{fig:1} we have plotted $\Delta_{\rm n}^{\rm d} F$ as function of the temperature for different values of the ``chemical potential like'' parameter $\mu$, while the other parameters are fixed. We see that for the three largest values of $\mu$ considered there is an interval in the middle where the native conformation is stable, while for low and high temperatures the denaturated protein is preferred. In other words, one has two unfolding transitions, a cold one and a warm one. This cold and warm unfolding seems to be a common feature of small globular proteins\cite{Privalov:90,Chen:89}. 

The smallest value $\mu_4 =1.74292\,...$ in Fig.\,\ref{fig:1} is a critical one, where the maximum of the stability function is at $\Delta_{\rm n}^{\rm d}F=0$, where the unfolded an folded states have equal probabilities. I.e., the lower curve of Fig.\ \ref{fig:2} has a maximum $0.5$. Qualitatively the parabolic plots in Fig.\,\ref{fig:1} corresponds well to experiments of Privalov {\it et al.}\cite{Privalov:86} on sperm whale metmyoglobin, where such conformational free energy differences were measured.       

The picture of cold and warm unfolding against different values of $\mu$ is substantiated by a glance at Fig.\,\ref{fig:2}, that shows the mean number of folded contacts relative to the system size given by 
\begin{equation}
   \label{n}
    n\equiv\frac{\sum_{i=0}^{N}i\, Z_i}{N\,\sum_{i=0}^{N} Z_i}
     =\frac{r}{N}\, \frac{N\,r^{N+1}-(N+1)\,r^N+1}{(1-r^{N+1})\,(1-r)} 
\end{equation}
For a complete denaturated protein $n=0$, while for a native one $n=1$.

It is now interesting to study the heat capacity as done in Fig.\,\ref{fig:3}. We obtain two peaks, which show both cold and warm unfolding. The peaks vanish as $\mu$ decreases towards its critical value. This feature is in accordance with the experiments of Privalov {\it et al.}\,\cite{Privalov:90,Privalov:86} on small globular proteins.

 By choosing $d=380$ K and $N_{\rm h}=67$ as the number of oscillators for $Z'$ (Eqs. \ref{eq:Z_h} and \ref{eq:Z'}), we see in Fig.\ \ref{fig:4} that the model is qualitatively in good correspondence with experimental data. We note the the specific choice  $d=380$K, corresponds to a wavelength $\sim 2\cdot 10^{-5}$m, which is in the IR region. 

Finally in this section it can be noted that the coupling between the water molecules (see Eq.\ \ref{eq:H_p}) resulting in the parameter $b$ did not have significant influence upon the qualitative behavior of our results. We have used a non-zero value of $b$, but replacing it with zero while adjusting other parameters would result in minor changes of the results obtained.

\section{Conclusion}
\label{sec:4}
We have studied a single-domain protein, which is supposed to follow a specific folding pathway. The chain-chain contact enthalpy and entropy increase linearly with the degree of folding. Each individual water molecule is modeled as a dipole in an external electrical field. Between the dipoles there are interactions that are incorporated in a mean field approximation. 

We find that the protein in the native form is in an intermediate temperature region, while it becomes denaturated for low and high temperatures. This cold and warm unfolding behavior is in accordance with experiments on small globular proteins\cite{Privalov:86,Chen:89,Privalov:90}, and is common to earlier models by Hansen {\it et al.}\cite{Hansen:98a,Hansen:99}, and Bakk {\it et al.}\cite{Bakk:00a,Bakk:00b,Bakk:00d}. 

By introducing vibrational modes, we find that the model also yields a more accurate quantitative representation of the heat capacity of proteins that undergo unfolding transitions at low and high temperatures\cite{Privalov:86,Privalov:90}.

%*************************************************************
% APPENDIX
%*************************************************************
\vspace{2cm}
\noindent
{\large\bf Appendix}
\begin{appendix}
\vspace{0.5cm}
\noindent
The assumed pair interaction $-\sum_jJ_{ij}\,{\bf s}_i\cdot{\bf s}_j$ between the dipole moment of water molecule $i$ and its neighbors $j$ is in a mean field consideration approximated by the term $-b\,{\bf m}\cdot{\bf s}_i$, where ${\bf s}_j$ is replaced by its average $<{\bf s}_j>={\bf m}$ and $b=\sum_j J_{ij}$. Such an approximation thus accounts to neglecting correlations between neighboring spins. The factor $b{\bf m}$ can now be regarded as an added electric field by which one obtains an effective (mean) electric field \cite{Hoye:80,Ma:85}
\begin{equation}
   \label{eq:E_e}
   E_{\rm e}=E+bm
\end{equation}
 that acts upon independent (or free) spins. However, when adding effective fields on all spins, interactions are counted twice which is compensated by an energy $1/2N_{\rm w}bm^2$ for a system counting of $N_{\rm w}$ water molecules. Thus, in a mean field treatment the pair interaction energy in Eq. \ref{eq:H_p} is approximated by
\begin{equation}
   \label{eq:H_p2}
   {\cal H}_{\rm p}\rightarrow -b{\bf m}\cdot\sum_i{\bf s}_i
                               +\frac{1}{2}N_{\rm w}bm^2.
\end{equation} 

 The partition function for one water molecule becomes
\begin{equation}
   \label{eq:Z_w}
   Z_{\rm w}=e^{-1/2\beta bm^2}Z_{\rm w}^{\rm e},
\end{equation}
$\beta\equiv 1/T$, with Boltzmanns constant $k_{\rm B}$ absorbed in T, and
\begin{equation}
    Z_{\rm w}^{\rm e}=2\pi\int_0^{\pi}d\vartheta\sin{\vartheta}\,
       e^{\beta E_e\cos{\vartheta}}
      =4\pi\frac{\sinh{\beta E_e}}{\beta E_e}. 
\end{equation}

The polarization (or ``magnetization'') $m$ is now obtained as\cite{Ma:85}
\begin{equation}
   \label{eq:m}
   m=\frac{\partial\ln{Z_{\rm w}^{\rm e}}}{\partial (\beta E_{\rm e})}
      =\coth{\beta{E}_{\rm e}}-\frac{1}{\beta{E}_{\rm e}}\quad ,
\end{equation} 
Inserting Eq.\ \ref{eq:E_e} in Eq.\ \ref{eq:m}, one obtain the relation between $m$ and $E$. 
\end{appendix}

%*************************************************************
% ACKNOWLEDGEMENTS
%*************************************************************
\vspace{2cm}
\noindent
{\large\bf Acknowledgements}

\vspace{0.5cm}
\noindent
We thank K. Sneppen for interesting and enlightening discussions. A. B. thanks the Research Council of Norway (Contract No. 129619/410) for financial support.

%*************************************************************
% LIST OF REFERENCES
%*************************************************************

%*************************************************************
% LIST OF FIGURE CAPTIONS
%*************************************************************
\vspace{2cm}
\noindent
{\large\bf Figure captions}

\vspace{0.5cm}
\noindent
{\bf Fig. 1. }Temperature dependence of the denaturated and native protein free energy difference ($\Delta_n^d F$ in Eq.\,\ref{eq:delta_f}) for different $\mu$. Other parameters according to Eq.\,\ref{eq:r} are $a=0.12$, $b=2$, and $M=10$. The absolute temperature is rescaled.

\vspace{0.5cm}
\noindent

\begin{figure}
   \caption{}
   \vspace{1cm}
   \centering{\epsfig{figure=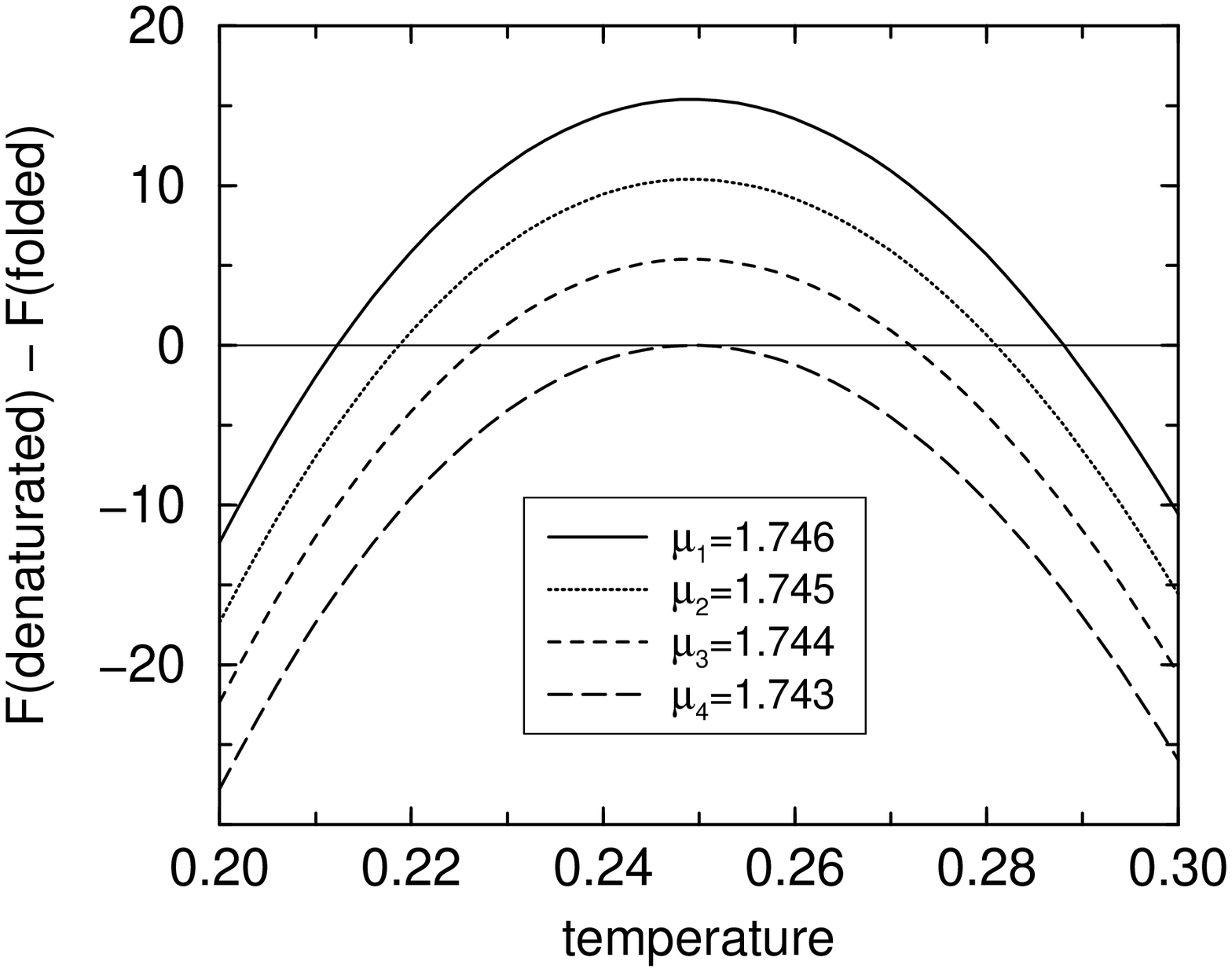,width=\linewidth}}
   \label{fig:1}
\end{figure}

\begin{figure}
   \caption{}
   \vspace{1cm}
   \centering{\epsfig{figure=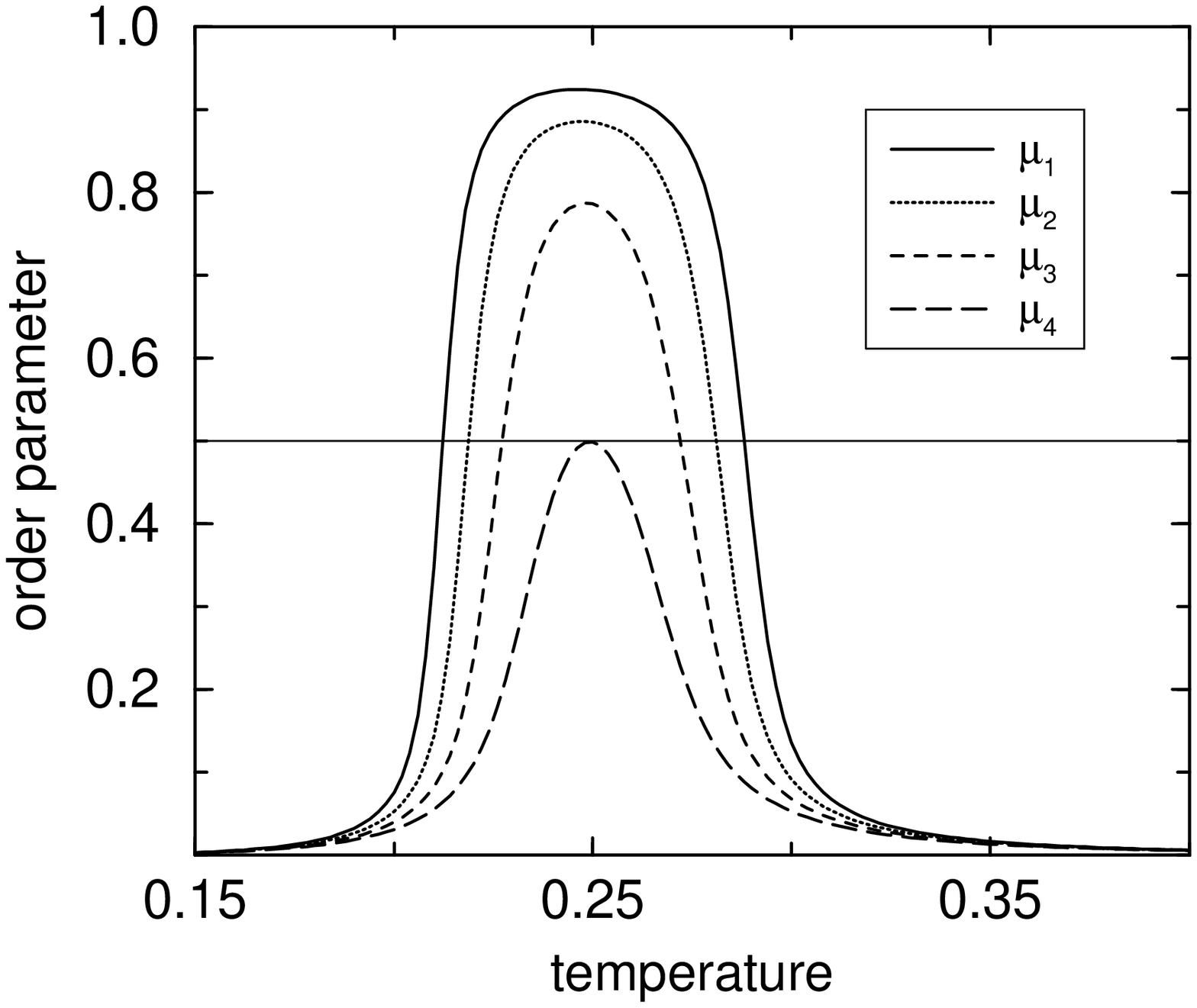,width=\linewidth}}
   \label{fig:2}
\end{figure}

\begin{figure}
   \caption{}
   \vspace{1cm}
   \centering{\epsfig{figure=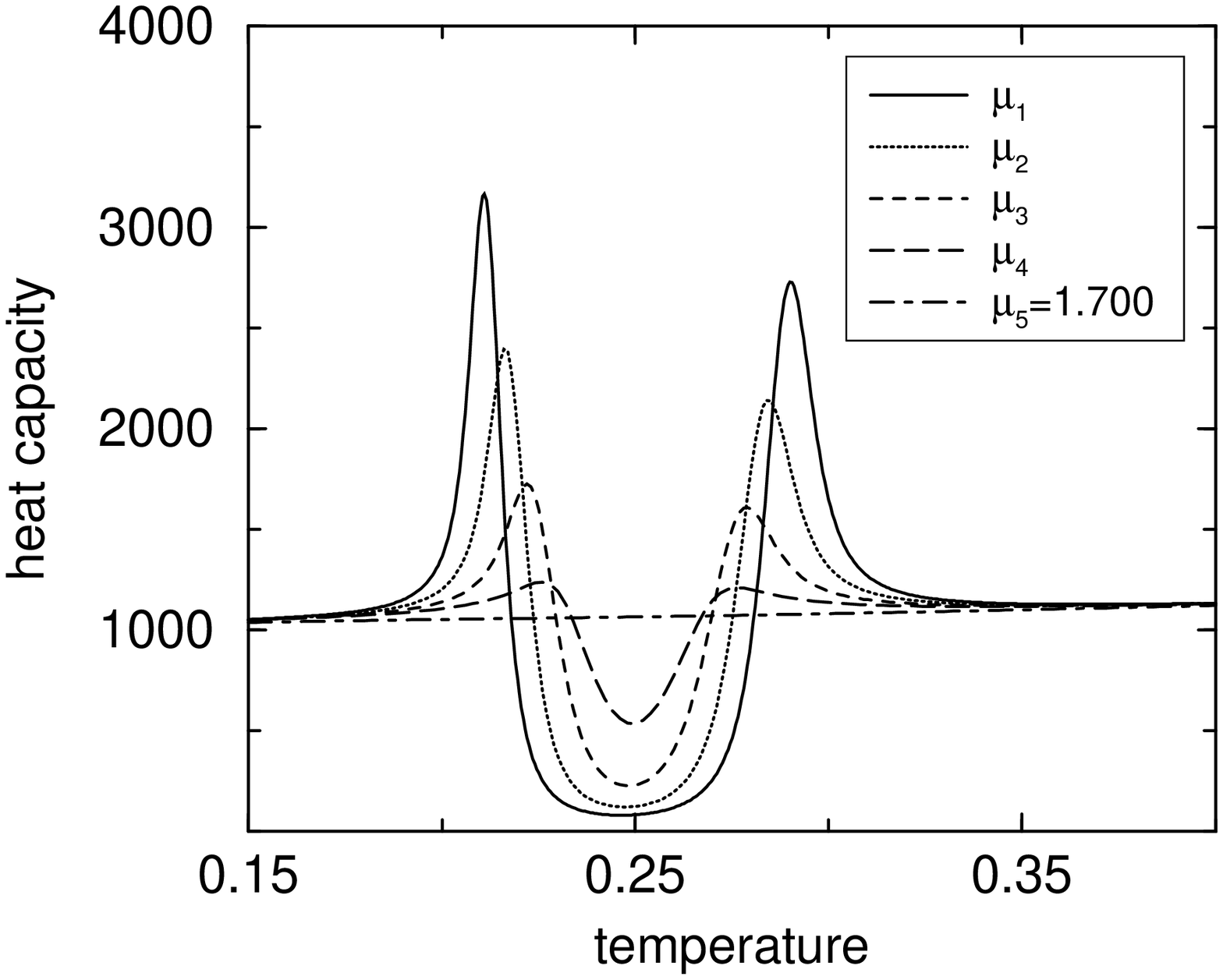,width=\linewidth}}
   \label{fig:3}
\end{figure}

\begin{figure}
   \caption{}
   \vspace{1cm}
   \centering{\epsfig{figure=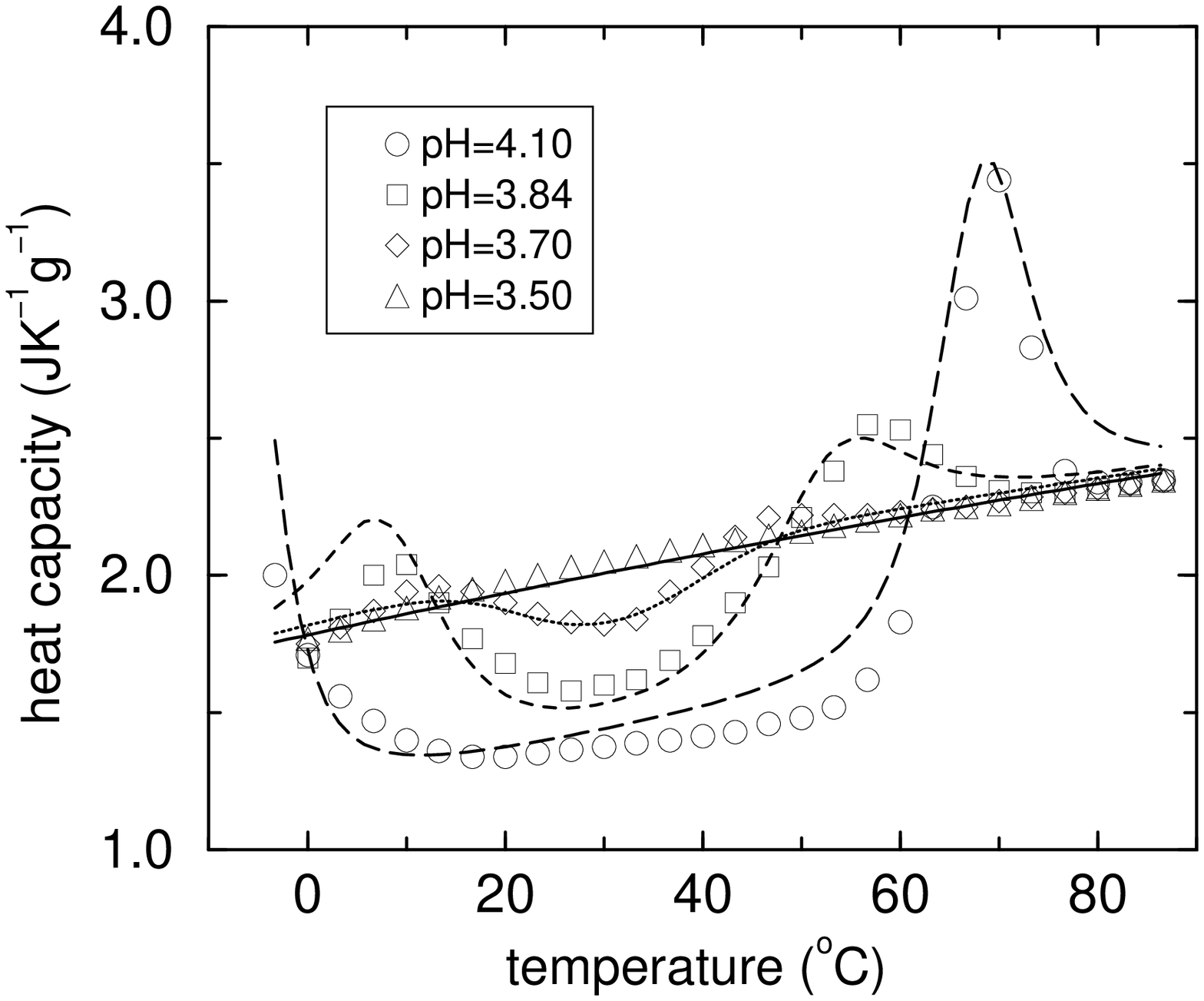,width=\linewidth}}
   \label{fig:4}
\end{figure}

\vspace{0.5cm}
\noindent
{\bf Fig. 2. }Temperature dependence of the order parameter $n$ for the corresponding parameter set as in Fig.\,\ref{fig:1}. Note that the maximum for $\mu_4$ is $n=0.5$, which corresponds to $\Delta_{\rm n}^{\rm d}F=0$ (see  Fig.\,\ref{fig:1}).

\vspace{0.5cm}
\noindent
{\bf Fig. 3. }Heat capacity for different $\mu$. Other parameters as in Fig.\,\ref{fig:1}. Note the smoothing of the peaks for decreasing $\mu$. The heat capacity for  $\mu_5$ is the hydration contribution of the denaturated protein.

\vspace{0.5cm}
\noindent
{\bf Fig. 4. }Heat capacity of metmyoglobin at different $\mu$ based upon $Z'$ in Eq.\ \ref{eq:Z'}. Experimental data from Privalov {\it et al.}\cite{Privalov:86} on metmyoglobin (Mb), where at pH $=3.50$ corresponds to a denaturated protein. At pH $=4.10$ the protein is folded between 20$^{\circ}$C and 50$^{\circ}$C, and has an unfolding transition at $\approx 70^{\circ}$C. One is also able to see from the data on pH $=4.10$ some destabilizing action at low temperatures.

%*************************************************************
% FIGURES
%*************************************************************
\newpage

\vspace*{1cm}
\centering{A. Bakk, J. S. H\o ye, and  A. Hansen}\\
\centering{\it Heat capacity of protein folding}\\

\newpage

\vspace*{1cm}
\centering{A. Bakk, J. S. H\o ye, and  A. Hansen}\\
\centering{\it Heat capacity of protein folding}\\

\newpage

\vspace*{1cm}
\centering{A. Bakk, J. S. H\o ye, and  A. Hansen}\\
\centering{\it Heat capacity of protein folding}\\

\newpage

\vspace*{1cm}
\centering{A. Bakk, J. S. H\o ye, and  A. Hansen}\\
\centering{\it Heat capacity of protein folding}\\


\vspace{2cm}
\begin{thebibliography}{99}

\bibitem{Creighton:92a} P. L. Privalov. Protein folding (ed. T. E. Creighton). W. H. Freeman and Company, New York, Ch. 3 (1992).

\bibitem{Anfinsen:73} C. B. Anfinsen. Principles that govern the folding of a protein chain (Nobel lecture). {\it Science}, {\bf 191}, 223-230 (1973).

\bibitem{Levinthal:68} C. Levinthal. Are there pathways for protein folding? {\it J. Chem. Phys.} {\bf 65}, 44-45 (1968).

\bibitem{Baldwin:99} R. L. Baldwin and G. D. Rose. Is protein folding hierarchic? I. Local structure and peptide folding. {\it Trends Biochem. Sci.} {\bf 24}, 26-33 (1999).

\bibitem{Wang:95} Yi Wang and D. Shortle. The equilibrium folding pathway of staphylococcal nuclease: identification of the most stable chain-chain interactions by NMR and CD spectroscopy. {\it Biochemistry US.} {\bf 34}, 15895-15905 (1995).

\bibitem{Hansen:98a} A. Hansen, M. H. Jensen, K. Sneppen, and G. Zocchi. Statistical mechanics of warm and cold unfolding in proteins. {\it Eur. Phys. J. B}, {\bf 6}, 157-161 (1998).

\bibitem{Hansen:99} A. Hansen, M. H. Jensen, K. Sneppen, and G. Zocchi. A model for the thermodynamics of globular proteins. {\it Physica A}, {\bf 270}, 278-287 (1999).

\bibitem{Hansen:98b} A. Hansen, M. H. Jensen, K. Sneppen, and G. Zocchi. A hierarchical scheme for cooperativity and folding in proteins. {\it Physica A}, {\bf 250}, 355-361 (1998).

\bibitem{Bakk:00a} A. Bakk. J. S. H\o ye, A. Hansen, Kim Sneppen, and M. H. Jensen. Pathways in two-state protein folding. {\it Biophys. J.} {\bf 79}, 2722-2727 (2000).

\bibitem{Bakk:00b} A. Bakk, J. S. H\o ye, A. Hansen, and Kim Sneppen. Thermodynamical implications of a protein model with water interactions. Submitted to Journal of Theoretical Biology (2000), e-print: cond-mat/0007078.

\bibitem{Creighton:92b} F. M. Richards. Protein folding (ed. T. E. Creighton). W. H. Freeman and Company, New York, Ch. 1 (1992).

\bibitem{Hoye:80} J. S. H\o ye and G. Stell. Statistical mechanics of polar fluids in electric fields. {\it J. Chem. Phys.} {\bf 72}, 1597-1613 (1980).

\bibitem{Ma:85} S.-K. Ma. Statistical mechanics. World Scientific, Philadelphia, USA. Ch. 27 (1985).

\bibitem{Privalov:86} P. L. Privalov, Yu. V. Griko, and S. Yu. Venyaminov. Cold denaturation of myoglobin. {\it J. Mol. Biol.} {\bf 190}, 487-498 (1986).

\bibitem{Plaxco:98} K.\ W.\ Plaxco, K.\ T.\ Simons, and D.\ Baker. Contact order, transition state placement and the refolding rates of single domain proteins. {\it J.\ Mol.\ Biol.} {\bf 277}, 985-994 (1998).

\bibitem{Leopold:92} P.\ E.\ Leopold, M.\ Montal, and J.\ N.\ Onuchic. Protein folding funnels - a kinetic approach to the sequence structure relationship. {\it P.\ Natl.\ Acad.\ Sci.\ USA}, {\bf 89}, 721-8725(1992). 

\bibitem{Galzitskaya:99} O.\ V.\ Galzitskaya and A.\ V.\ Finkelstein. A theoretical search for folding/unfolding nuclei in three-dimensional protein structures. {\it Proc.\ Natl.\ Acad.\ Sci.\ USA.} {\bf 96}, 11299-11304 (1999).

\bibitem{Bryngelson:87} J.\ D.\ Bryngelson and P.\ G.\ Wolynes. Spin-glasses and the statistical-mechanics of protein folding. {\it Proc.\ Natl.\ Acad.\ Sci.\ USA.} {\bf 84}, 7524-7528 (1987).

\bibitem{Zwanzig:95} R.\ Zwanzig. Simple model of protein folding kinetics. Proc.\ Natl.\ Acad.\ Sci.\ USA, {\bf 92}, 9801-9804 (1995).

\bibitem{Edsall:35} J. T. Edsall. Apparent molal heat capacities of amino acids and other organic compounds. {\it J. Am. Chem. Soc.} {\bf 57}, 1506-1507 (1935).

\bibitem{Privalov:92} P. L. Privalov and G. I. Makhatadze. Contribution of hydration and non-covalent interactions to the heat capacity effect on protein unfolding. {\it J. Mol. Biol.} {\bf 224}, 715-723 (1992).

\bibitem{Privalov:88} P. L. Privalov and S. J. Gill. Stability of protein structure and hydrophobic interaction. {\it Adv. Protein Chem.} {\bf 39}, 191-234 (1988).

\bibitem{Frank:45} H. S. Frank and M. W. Evans. Free volume and entropy in condensed systems. III Entropy in binary liquid mixtures; partial molal entropy in dilute solutions; structure and thermodynamics in aqueous electrolytes. {\it J. Chem. Phys.} {\bf 13}, 507-532 (1945).

\bibitem{Bakk:00c} A.\ Bakk and J.\ S.\ H\o ye. A microscopic argument for the anomalous hydration heat capacity increment upon solvation of apolar substances. Submitted to {\it J.\ Chem.\ Phys.}

\bibitem{Privalov:90} P. L. Privalov. Cold denaturation of proteins. {\it Crit. Rev. Biochem. Mol.} {\bf 25}, 281-305 (1990).

\bibitem{Chen:89} B. Chen and J. A. Schellman. Low-temperature unfolding of a mutant of phage T4 lysozyme. 1. Equilibrium Studies. {\it Biochemistry US.} {\bf 28}, 685-691 (1989).

\bibitem{Bakk:00d} A. Bakk, A. Hansen, and Kim Sneppen. A protein model exhibiting three folding transitions. To appear in {\it Physica A}.

\end{thebibliography}
\end{document}